\documentstyle[preprint,aps]{revtex}

\def\Rey{{\text\it Re}}          
\def\hatx{\hat{\bf x}}           
\def\lesssim{\,
  \mbox{\raisebox{-0.7ex}{\mbox{$\stackrel{\textstyle<}{\sim}$}}}\;}
\def\bcdot{\,{\bf\cdot}\,}
\def\bnabla{{\bf\nabla}}
\def\etal{\mbox{\it et al.}}
\def\argxt{\!\left({\bf x}, t\right)}
\def\argx{\!\left({\bf x}\right)}
\def\argz{\!\left(z\right)}
\def\argphi{\!\left\{\phi\right\}}
\def\argRe{\!\left(\Rey\,\right)}
\def\argzeta{\!\left(\zeta\right)}

\def\ceps{c_{\varepsilon}}
\def\cbar{\overline{c_{\varepsilon}}}
\def\Rti{\Rey\to\infty}
\def\clow{\underline{c_{\varepsilon}}}
\def\RES{\Rey_{ES}}
\def\RBi{\Rey_{B}}
\def\kES{k_{ES}}
\def\Rzpk{R_0\argphi\!\left({\bf k}\right)}
\tighten
\title{Variational bound on energy dissipation in plane Couette flow}
\author{Rolf Nicodemus\footnote{Electronic address:
	    nicodemu@mailer.uni-marburg.de},
        Siegfried Grossmann\footnote{Electronic address:
	    grossmann\_s@stat.physik.uni-marburg.de},
	and Martin Holthaus\footnote{Electronic address:
	    holthaus@stat.physik.uni-marburg.de}}
\address{Fachbereich Physik der Philipps-Universit\"at, \\
	Renthof 6, D--35032 Marburg, Germany}
\date{\today}
\narrowtext
\begin{document}
\maketitle
\begin{abstract}
We present numerical solutions to the extended Doering--Constantin
variational principle for upper bounds on the energy dissipation rate
in turbulent plane Couette flow. Using the compound matrix technique
in order to reformulate this principle's spectral constraint, we derive
a system of equations that is amenable to numerical treatment in the
entire range from low to asymptotically high Reynolds numbers. Our
variational bound exhibits a minimum at intermediate Reynolds numbers,
and reproduces the Busse bound in the asymptotic regime. As a consequence
of a bifurcation of the minimizing wavenumbers, there exist two length
scales that determine the optimal upper bound: the effective width of the
variational profile's boundary segments, and the extension of their flat
interior part.
\end{abstract}


\section{Introduction}

Can one derive upper bounds on the rate of energy dissipation in turbulent
shear flows that are both mathematically rigorous and physically meaningful,
i.e.\ ``sharp''? This question is of enormous conceptual importance in
the theory of turbulence, since the energy dissipation rate is the central
object in classical turbulence theories~\cite{K41,O41,H48,W48,O45}, and its
behavior at high Reynolds numbers may reveal whether there are
intermittency corrections to classical scaling~\cite{G95,SS95}. Since full
numerical simulations of shear flows with Reynolds numbers of the order of
$10^6$ or above are out of reach, and will remain so in the foreseeable
future, rigorous estimates of the dissipation rate derived directly from the
Navier--Stokes equations are one of the few tools left to the theorist for
approaching such questions.

The idea of bounding the rate of energy dissipation with the help
of rigorous inequalities, rather than trying to solve the equations of
motion under plausible, but essentially uncontrolled approximations, saw
its first prime in the early seventies, with the formulation and further
exploration of the Howard--Busse theory~\cite{H72}. In particular,
Busse~\cite{B70,B78,B96} was able to derive an asymptotic upper bound
on energy dissipation in plane Couette flow that is in accordance with
classical theories, but lies roughly an order of magnitude above
experimentally measured data~\cite{R59,LFS92}.

After this state of affairs had remained unchanged for about 25 years,
Doering and Constantin put forward a new variational principle for
computing quantities characterizing turbulent flows~\cite{DC94,CD95,DC96}.
Whereas the first applications of this principle to the plane Couette
problem~\cite{DC94,DC92,GGHL95} did not exhaust the principle and produced
bounds on the dissipation rate that could not compete with Busse's,
subsequent analysis by Kerswell~\cite{K97} revealed that the
Doering--Constantin principle, after being extended along the lines worked
out in Ref.~\cite{NGH97a}, actually reproduces the Busse bound
in the asymptotic regime. Kerswell arrived at this conclusion by mapping
the extended Doering--Constantin principle to another variational principle
that also occurs in Busse's theory, and then applying Busse's method
of solution. However, as emphasized by Kerswell himself~\cite{K97}, there is
still no mathematical proof that Busse's solution is really extremal, although
this is generally believed to be the case. Moreover, the Busse bound
pertains to the limit of asymptotically high Reynolds numbers only.
Hence, what is needed is a solution to the extended Doering--Constantin
principle that (i) fully exhausts this principle, (ii) spans the entire range
from low to asymptotically high Reynolds numbers, and (iii) is logically
independent from Busse's solution.

In this paper, we present a full-fledged numerical analysis of the improved
Doering--Constantin variational principle for the plane Couette flow that
meets these requirements. The goal here is not merely to derive the best
possible bound, but we also wish to explore the working principles of the new
method. Even if statements concerning the existence of non-classical scaling
remain out of reach at present, the potential power of the new variational
principle lies in its great conceptual clarity. Once thoroughly understood, it
might serve as a germ for the formulation of a still more advanced principle
that could actually allow one to attack the scaling questions.

We organize our material as follows: after having stated the variational
principle in Sec.~\ref{S_VP}, Sec.~\ref{S_SC} will be devoted to the
resolution of the most demanding technical difficulty, the implementation of
the principle's spectral constraint. It is this constraint that seems to have
hindered previous numerical work~\cite{DH97} to reach the asymptotic
regime. Our results will then be described in Sec.~\ref{S_RE}; the final
Sec.~\ref{S_CO} offers a concluding discussion. Some important technical
details concerning the derivation of the system of differential equations
actually used in the numerical computations can be found in
Appendix~\ref{A_1}, while Appendix~\ref{A_2} contains a brief, but
instructive example that illustrates the key points of our method, and
underlines its accuracy, without too much technical burden.

Since the numerical work is fairly involved, it would hardly have been
possible to carry through the entire investigation without some analytical
guidance. Our guide is the example problem of the Couette flow without
spanwise degrees of freedom, referred to as the {\em restricted Couette
problem\/} in the following, which has been treated in detail in
Ref.~\cite{NGH97bc}, and which shows several features that will reappear here
in the context of the unrestricted problem.

\section{The variational principle}
\label{S_VP}

We consider a layer of an incompressible fluid with kinematic viscosity $\nu$
that is confined between two infinitely extended, parallel rigid plates
separated by a distance $h$. The lower plate, coinciding with the plane
$z=0$ of a Cartesian coordinate system, is fixed, whereas the upper plate
at $z = h$ is sheared with constant velocity $U$ in the positive
$x$-direction. The dynamics of the fluid's velocity field ${\bf u}\argxt$
are determined by the equations
\begin{eqnarray}
  \partial_t {\bf u} + {\bf u}\bcdot\bnabla{\bf u} +\bnabla p =
    \nu\,\Delta{\bf u} & \quad & \mbox{(Navier--Stokes equations)},
  \label{2NSG}\\
  \bnabla\bcdot{\bf u} = 0 & \quad & \mbox{(incompressibility)},
  \label{2INC} \\
  {\bf u}\!\left(x,y,0,t\right) = {\bf 0}, \quad
    {\bf u}\!\left(x,y,h,t\right) = U\hatx & \quad &
  \mbox{(no-slip boundary conditions)}; \label{2NSB}
\end{eqnarray}
here $p$ denotes the kinematic pressure and $\hatx$ is the unit vector in
$x$-direction. In addition, in both $x$- and $y$-direction periodic
boundary conditions (b.c.) are imposed on ${\bf u}$ and $p$.
The time-averaged rate of dissipated energy per mass is given by
\begin{equation}
  \varepsilon_T\equiv\frac{1}{T}\int\limits_{\!0}^{\;T}\!{\rm d}t
  \left\{\frac{\nu}{\Omega}\int\limits_{\Omega}{\rm d}^3\!x
  \left[\sum_{i,j=x,y,z}\!\!\left(\partial_j u_i\right)^2\right]\right\},
  \label{2EDR}
\end{equation}
where $\Omega$ is the periodicity volume. Our aim is to derive a
mathematically rigorous upper bound on the long-time limit $\varepsilon$
of $\varepsilon_T$,
\begin{equation}
  \varepsilon\equiv\lim_{T\to\infty}\varepsilon_T.
\end{equation}
Equivalently, we consider the non-dimensionalized dissipation rate
\begin{equation}
  \ceps\argRe\equiv\frac{\varepsilon}{U^3 h^{-1}},
  \label{2NDD}
\end{equation}
where $\Rey=Uh/\nu$ is the Reynolds number. For calculating the bound
we will employ the {\em background flow method\/} as recently put forward
by Doering and Constantin~\cite{DC94,CD95,DC96,DC92}, and developed further
in Ref.~\cite{NGH97a}.

It is known~\cite{DC94} that the rate of energy dissipated by the
laminar flow ${\bf u}\argxt=\left(Uz/h\right)\hatx$ provides a rigorous {\em
lower\/} bound $\clow$ on $\ceps$ for all $\Rey$, namely
$\ceps\argRe\ge\clow\argRe\equiv\Rey^{-1}$. Moreover, energy stability
theory~\cite{J76,DR81} guarantees that the laminar flow is the only possible
stable flow at least up to the energy stability limit $\RES$. Hence, we have
the exact identity
\begin{equation}
  \ceps\argRe=\Rey^{-1}\quad\mbox{for}\quad\Rey<\RES\approx 82.65.
  \label{2EST}
\end{equation}
From an early generalization of energy stability theory, the Optimum Theory
devised by Busse~\cite{B70,B78,B96}, one obtains an approximate
{\em asymptotic\/} upper bound~\cite{note}:
\begin{equation}
  \lim_{\Rti}\ceps\argRe\lesssim 0.010.
  \label{2CBU}
\end{equation}
Note that the right hand side of this inequality does not depend on the
Reynolds number, so that this bound is in accordance with classical turbulence
theories~\cite{K41,O41,H48,W48,O45}. (For a discussion of the connection
between the $\Rey$-dependence of $\ceps\argRe$ and possible intermittency
corrections, see Ref.~\cite{G95}.)


The background flow approach can likewise be considered as a generalization of
energy stability theory, but it has a distinctly different twist. Instead of
decomposing the velocity field ${\bf u}$ into a $z$-dependent,
plane-averaged mean flow and the fluctuations around this flow, as Busse did
in his Optimum Theory, Doering and Constantin~\cite{DC92} take up an idea
by Hopf~\cite{H41} and write
\begin{equation}
  {\bf u}\argxt = {\bf U}\argx + {\bf v}\argxt. \label{2DEC}
\end{equation}
Now the stationary and divergence-free auxiliary field ${\bf U}\argx$,
dubbed the background flow, has to carry the physical b.c., namely
${\bf U}\!\left(x,y,0\right)={\bf 0}$, ${\bf U}\!\left(x,y,h\right)=U\hatx$,
and ${\bf U}\argx$ is periodic in $x$- and $y$-direction. But apart from
these fairly mild specifications, the background flow remains completely
arbitrary. Hence, this decomposition~(\ref{2DEC}) serves as the starting
point of a variational principle~\cite{DC94}: plug it into the
expression~(\ref{2EDR}) for the energy dissipation rate, get rid of the
deviations ${\bf v}\argxt$ with the help of both the equations of motion
(\ref{2NSG})--(\ref{2NSB}) and rigorous inequalities that bound the
dissipation rate from above, and then adjust the background flow such that
this upper bound becomes as low as possible.

For carrying through this program, we restrict ourselves to background flows
that are given by a merely height-dependent profile $\phi\argzeta$, which
should reflect the symmetry of the Couette geometry. That is, we admit only
background flows of the form
\begin{equation}
  {\bf U}\argx\equiv U\phi\argzeta\hatx,
\end{equation}
with
\begin{equation}
  \phi\!\left(0\right)=0,\quad\phi\!\left(1\right)=1,
  \quad\phi\argzeta=1-\phi\!\left(1-\zeta\right), \label{2PRO}
\end{equation}
where $\zeta\equiv z/h$ is the dimensionless coordinate in cross-stream
direction. The resulting inequality that bounds the dissipation rate
then reads~\cite{NGH97a}
\begin{equation}
  \ceps\argRe\leq \left[1+\frac{a^2}{4\left(a-1\right)}\,
    D\argphi\right]\Rey^{-1}, \label{2VAP}
\end{equation}
where $a>1$ is a dimensionless balance parameter, and $D\argphi$ denotes the
profile functional
\begin{equation}
  D\argphi\equiv\int\limits_{\!0}^{\;1}\!{\rm d}\zeta
  \left[\phi^{\prime}\argzeta\right]^2-1. \label{2FUN}
\end{equation}
The bound on the dissipation rate provided by the r.h.s.\ of the
inequality~(\ref{2VAP}) can be minimized by varying both the profile $\phi$
and the balance parameter $a$. However, there is a substantial technical
difficulty: {\em the inequality (\ref{2VAP}) is valid only as long as the
profile $\phi$ satisfies a spectral constraint\/}~\cite{DC94}. This spectral
constraint demands that all eigenvalues $\lambda$ of the linear eigenvalue
problem
\begin{eqnarray}
  \lambda {\bf V} & = & - 2\,h^2\Delta{\bf V} + R\,\phi^{\prime}
    \left(\begin{array}{ccc} 0 & 0 & 1 \\ 0 & 0 & 0 \\ 1 & 0 & 0
      \end{array}\right)
    {\bf V} + \bnabla P, \nonumber \\
  0 & = & \bnabla\bcdot{\bf V}, \quad
    {\bf V}\!\left(\partial\Omega\right) = {\bf 0}
  \label{2SPC}
\end{eqnarray}
for the stationary velocity fields ${\bf V}$ be positive. Since the
background flow already carries the physical boundary conditions, ${\bf V}$
has to satisfy homogeneous b.c., that is, ${\bf V}$ has to vanish on the
material walls $\partial\Omega$ at $z=0$ and $z=h$, and to be periodic in
$x$- and $y$-direction. Note that the balance parameter $a$ enters into this
eigenvalue problem through the rescaled Reynolds number
\begin{equation}
  R\equiv\frac{a}{a-1}\,\Rey. \label{2RES}
\end{equation}

In effect, the spectral constraint~(\ref{2SPC}) restricts, for each Reynolds
number, the combinations of profiles and balance parameters that are
admissible to the variational principle~(\ref{2VAP}). The highly non-trivial
task of evaluating this constraint consists in determining, for each
candidate profile $\phi$ separately, that rescaled Reynolds number $R$ where
the smallest eigenvalue passes through zero. We will denote this number as
$R_c\argphi$ in the following.

{\em Provided\/} we know how to compute $R_c\argphi$ --- this task will be
taken up in the following section ---, each $\phi$ leads to a rigorous upper
bound on $\ceps$ in the interval $0\leq\Rey<R_c\argphi$. After optimization
of the balance parameter, this bounds adopts the form~\cite{NGH97a}
\begin{equation}
  \ceps\leq\left\{
    \begin{array}{lll}
      \left[1+D\argphi\right]\Rey^{-1} &
      \mbox{for} &
      0\leq\Rey<\frac{1}{2}R_c\argphi \\
      \left[1+
      \frac{D\argphi R_c\argphi^2}
        {4\left(R_c\argphi-\Rey\right)\Rey}\right]\Rey^{-1} &
      \mbox{for} &
      \frac{1}{2}R_c\argphi\leq\Rey<R_c\argphi
    \end{array}\right..
  \label{2CEP}
\end{equation}
Figure~\ref{F_1} shows the graph of such an upper bound produced by a generic
profile $\phi$. When a suitable class of test profiles has been selected, the
lower envelope of all graphs stemming from the individual profiles then is the
optimal upper bound on $\ceps\argRe$ that can be obtained from this class,
i.e., the solution to the variational principle~(\ref{2VAP}) within this
class.

\section{Evaluation of the spectral constraint}
\label{S_SC}

We non-dimensionalize the problem by choosing the gapwidth $h$ as the unit of
length. By virtue of the linearity of the equations~(\ref{2SPC}), and the
homogeneous b.c.\ imposed on the eigenvectors ${\bf V}$, it is not necessary
to specify a scale of velocity explicitly. For the sake of notational
simplicity we denote all non-dimensionalized quantities by the same symbols as
their dimension-carrying counterparts. For instance, the velocity field
${\bf V}\argx$ now is regarded as a dimensionless vector function of the
dimensionless coordinates $x$, $y$ and $z$.

Utilizing the periodic b.c., we start from the ansatz
\[
  {\bf V}\argx\equiv{\bf v}\argz {\rm e}^{{\rm i}\left(k_x x+k_y y\right)},
    \quad P\argx\equiv p\argz {\rm e}^{{\rm i}\left(k_x x+k_y y\right)}
\]
and transform the eigenvalue equations~(\ref{2SPC}) into the system
\begin{eqnarray}
  \lambda v_x & = &
    - 2 \left(\partial_z^2 - k^2\right) v_x +
    R\,\phi^{\prime} v_z + {\rm i}\,k_x p,
  \label{3VX1} \\
  \lambda v_y & = &
    - 2 \left(\partial_z^2 - k^2\right) v_y +
    {\rm i}\,k_y p,
  \label{3VY1} \\
  \lambda v_z & = &
    - 2 \left(\partial_z^2 - k^2\right) v_z +
    R\,\phi^{\prime} v_x + p^{\prime},
  \label{3VZ1} \\
  0 & = & {\rm i}\,k_x v_x + {\rm i}\,k_y v_y + v_z^{\prime},
  \label{3DIV}
\end{eqnarray}
with
\[
  k\equiv\sqrt{k_x^2+k_y^2}.
\]
The corresponding b.c.\ for the function ${\bf v}\argz$ read
\begin{equation}
  {\bf v}\!\left(0\right) = {\bf v}\!\left(1\right) = {\bf 0}.
  \label{3BCN}
\end{equation}
If $k_y=0$ we encounter the model problem of a Couette flow without
spanwise degrees of freedom. Since this restricted problem has already
been studied in detail in Ref.~\cite{NGH97bc}, we only need to consider
$k_y\neq 0$ here.

Defining the linear operator
\[
  {\cal L}\equiv 2 \left(\partial_z^2 - k^2\right) + \lambda,
\]
Eqs.~(\ref{3VX1}) and (\ref{3VY1}) together with Eq.~(\ref{3DIV}) yield
\begin{equation}
  p=\frac{1}{k^2}\left[{\cal L}v_z^{\prime} + {\rm i}\,k_x R\,\phi^{\prime}
  v_z\right]. \label{3PRE}
\end{equation}
Hence, Eq.~(\ref{3VX1}) can be brought into the form
\begin{equation}
  {\cal L}v_x=\frac{1}{k^2}\left[k_y^2 R\,\phi^{\prime}v_z +
  {\rm i}\,k_x{\cal L}v_z^{\prime}\right], \label{3VX2}
\end{equation}
and we finally obtain from Eq.~(\ref{3VZ1}) a rather intricate equation
for $v_z\argz$:
\begin{eqnarray}
  {\cal L}^2\left(\partial_z^2-k^2\right)v_z -
  4\,\frac{\phi^{(2)}}{\phi^{\prime}}\,
    {\cal L}\left(\partial_z^2-k^2\right)v_z^{\prime} -
  2\left[\frac{\phi^{(3)}}{\phi^{\prime}}-
    2\left(\frac{\phi^{(2)}}{\phi^{\prime}}\right)^2\right]{\cal L}
   \left(\partial_z^2-k^2\right)v_z & & \nonumber \\
  +\quad\left(k_y R\,\phi^{\prime}\right)^2v_z + {\rm i}\,k_x R\,\phi^{\prime}
    \left\{
    2{\cal L}v_z^{\prime}+\frac{\phi^{(2)}}{\phi^{\prime}}{\cal L}v_z+
    4\left[\frac{\phi^{(3)}}{\phi^{\prime}}-
      \left(\frac{\phi^{(2)}}{\phi^{\prime}}\right)^2\right]v_z^{\prime}
  \right. & & \nonumber \\
    \left. +\quad 2\left[\frac{\phi^{(4)}}{\phi^{\prime}}-
      3\,\frac{\phi^{(3)}\phi^{(2)}}{\left(\phi^{\prime}\right)^2}+
      2\left(\frac{\phi^{(2)}}{\phi^{\prime}}\right)^3\right]v_z\right\} &
      = & 0,
  \label{3SL1}
\end{eqnarray}
where we rely on the assumption that $\phi^{\prime}\argz$ does not vanish for
all $z$ in the interval $\left[0,1\right]$. The corresponding b.c.\ for $v_z$
are given by
\begin{equation}
  v_z\argz|_{z=0,1}=0,\quad v_z^{\prime}\argz|_{z=0,1}=0, \quad\mbox{and}\quad
  {\cal L}\left(\partial_z^2-k^2\right)v_z\argz|_{z=0,1}=0. \label{3SL2}
\end{equation}
Equations~(\ref{3SL1}) and (\ref{3SL2}) constitute a sixth-order
Sturm--Liouville eigenvalue problem. If one succeeds in solving this problem,
that is, in computing an eigenvalue $\lambda$ and its eigenfunction
$v_z^{(\lambda)}\argz$ for some fixed profile $\phi$, wavenumbers $k_x$
and $k_y$, and a rescaled Reynolds number $R$, then one can determine the
associated components $v_x^{(\lambda)}\argz$ and $v_y^{(\lambda)}\argz$,
together with $p^{(\lambda)}\argz$, from Eqs.~(\ref{3VX2}), (\ref{3VY1}), and
(\ref{3PRE}), so that an eigenvalue and a corresponding eigenvector
$\left({\bf V}^{(\lambda)}\argx,P^{(\lambda)}\argx\right)$ of the original
eigenvalue problem (\ref{2SPC}) have been found.

Since we cannot hope to find an analytical solution to this formidable
problem for each candidate profile $\phi$, we have to resort to a numerical
treatment. This requires to bring the Sturm--Liouville problem into a form
that can actually be dealt with on a computer. We defer the technical details
of this reformulation, that rely heavily on the so-called compound
matrix method~\cite{S92}, to Appendix~\ref{A_1}, and present only the final
equations here. Although these equations may appear, on a cursory glance,
even more cumbersome than the original problem posed by Eqs.~(\ref{3SL1}) and
(\ref{3SL2}), they have actually been found well suited for computing the
``critical'' rescaled Reynolds numbers $R_c\argphi$ for all situations of
practical interest, as will be amply demonstrated in the following section.

To begin with, we define the abbreviations
\begin{equation}
  R_x\equiv\frac{k_x R}{k^2},\quad R_y\equiv\frac{k_y R}{k^2}, \label{3ABB}
\end{equation}
and the auxiliary functions
\begin{eqnarray}
  \phi_1\argz & \equiv & \phi^{\prime}\argz, \nonumber \\
  \phi_2\argz & \equiv & \frac{\phi^{(2)}\argz}{2\,k}, \nonumber \\
  f_1\argz    & \equiv & 2\,\frac{\phi^{(2)}\argz}{\phi^{\prime}\argz},
    \nonumber \\
  f_2\argz    & \equiv &
    3\,k+\frac{2}{k}\,\frac{\phi^{(3)}\argz}{\phi^{\prime}\argz}-
    \frac{4}{k}\left(\frac{\phi^{(2)}\argz}{\phi^{\prime}\argz}\right)^2,
    \nonumber \\
  g_1\argz    & \equiv & \phi^{\prime}\argz-\frac{1}{k^2}\,\phi^{(3)}\argz+
    \frac{1}{k^2}\,\frac{\left(\phi^{(2)}\argz\right)^2}{\phi^{\prime}\argz},
    \nonumber \\
  g_2\argz    & \equiv & \frac{1}{2\,k}\,\phi^{(2)}\argz-
    \frac{1}{2\,k^3}\,\phi^{(4)}+
    \frac{3}{2\,k^3}\,
    \frac{\phi^{(3)}\argz\phi^{(2)}\argz}{\phi^{\prime}\argz}-
    \frac{1}{k^3}\,\frac{\left(\phi^{(2)}\argz\right)^3}
      {\left(\phi^{\prime}\argz\right)^2}, \nonumber \\
  F_{R_y}\argz & = &
    k + \frac{1}{k}\,\frac{\phi^{(3)}\argz}{\phi^{\prime}\argz}-
    \frac{2}{k}\left(\frac{\phi^{(2)}\argz}{\phi^{\prime}\argz}\right)^2-
    \frac{1}{4\,k}\left(R_y\phi^{\prime}\argz\right)^2. \nonumber \\
  & & \label{3AUX}
\end{eqnarray}
The compound matrix technique outlined in Appendix~\ref{A_1} then yields the
following system of 20 ordinary, first-order complex differential equations:
\begin{eqnarray}
  y_1^{\prime} & = & k\left[-3\,y_1+y_2-{\textstyle\frac{1}{2}}\,y_4-
      {\textstyle\frac{1}{2}}\,y_6\right], \nonumber \\
  y_2^{\prime} & = &
    k\left[-3\,y_2+y_3+y_5-{\textstyle\frac{1}{2}}\,y_8\right],
    \nonumber \\
  y_3^{\prime} & = & k\left[-3\,y_3+y_4+y_6\right], \nonumber \\
  y_4^{\prime} & = & k\left[{\textstyle\frac{3}{2}}\,y_3-3\,y_4+y_7+
      {\textstyle\frac{1}{2}}\,y_{10}\right] +
    f_1\left[-2\,y_2+y_4\right] - f_2 y_1
    - {\rm i}\,R_x\left[\phi_1 y_2+
      \phi_2\left(y_1+{\textstyle\frac{1}{2}}\,y_3\right)\right], \nonumber \\
  y_5^{\prime} & = & k\left[-3\,y_5+y_6+{\textstyle\frac{1}{2}}\,y_9+
      y_{11}\right], \nonumber \\
  y_6^{\prime} & = & k\left[-3\,y_6+y_7+y_8+{\textstyle\frac{1}{2}}\,y_{10}+
      y_{12}\right], \nonumber \\
  y_7^{\prime} & = &
    k\left[{\textstyle\frac{3}{2}}\,y_6-3\,y_7+y_9+y_{13}\right] +
    f_1\left[-y_1-2\,y_5+y_7\right]
    - {\rm i}\,R_x\left[\phi_1 y_5+{\textstyle\frac{1}{2}}\,\phi_2 y_6+
      g_1y_1\right], \nonumber \\
  y_8^{\prime} & = & k\left[-3\,y_8+y_9+y_{14}\right], \nonumber \\
  y_9^{\prime} & = & k\left[{\textstyle\frac{3}{2}}\,y_8-3\,y_9+y_{10}+
      y_{15}\right] +
    f_1\left[-y_2+y_9\right] + f_2 y_5
    - {\rm i}\,R_x\left[\phi_2\left(-y_5+
      {\textstyle\frac{1}{2}}\,y_8\right)+g_1 y_2\right], \nonumber \\
  y_{10}^{\prime} & = & k\left[-3\,y_{10}+y_{16}\right] +
    f_1\left[-y_3+2\,y_8+y_{10}\right] + f_2 y_6
    - {\rm i}\,R_x\left[-\phi_1 y_8-\phi_2 y_6+g_1 y_3\right], \nonumber \\
  y_{11}^{\prime} & = &
    k\left[-3\,y_{11}+y_{12}+{\textstyle\frac{1}{2}}\,y_{15}+
      {\textstyle\frac{1}{2}}\,y_{17}\right], \nonumber \\
  y_{12}^{\prime} & = &
    k\left[-3\,y_{12}+y_{13}+y_{14}+{\textstyle\frac{1}{2}}\,y_{16}\right],
    \nonumber \\
  y_{13}^{\prime} & = &
    k\left[{\textstyle\frac{3}{2}}\,y_{12}-3\,y_{13}+y_{15}-
      {\textstyle\frac{1}{2}}\,y_{19}\right] +
    f_1\left[-2\,y_{11}+y_{13}\right] + F_{R_y} y_1 \nonumber \\
  & + & {\rm i}\,R_x\left[-\phi_1 y_{11}-
      {\textstyle\frac{1}{2}}\,\phi_2 y_{12}+g_2 y_1\right], \nonumber \\
  y_{14}^{\prime} & = & k\left[-3\,y_{14}+y_{15}+y_{17}\right], \nonumber \\
  y_{15}^{\prime} & = &
    k\left[{\textstyle\frac{3}{2}}\,y_{14}-3\,y_{15}+y_{16}+y_{18}-
      {\textstyle\frac{1}{2}}\,y_{20}\right] +
    f_1 y_{15} + f_2 y_{11} + F_{R_y} y_2 \nonumber \\
  & + & {\rm i}\,R_x\left[\phi_2\left(y_{11}-
      {\textstyle\frac{1}{2}}\,y_{14}\right) + g_2 y_2\right], \nonumber \\
  y_{16}^{\prime} & = &
    k\left[-3\,y_{16}+y_{19}\right] + f_1\left[2\,y_{14}+y_{16}\right] +
    f_2 y_{12} + F_{R_y} y_3
    + {\rm i}\,R_x\left[\phi_1 y_{14}+\phi_2 y_{12}+g_2 y_3\right],
    \nonumber \\
  y_{17}^{\prime} & = & k\left[-3\,y_{17}+y_{18}-
      {\textstyle\frac{1}{2}}\,y_{20}\right], \nonumber \\
  y_{18}^{\prime} & = &
    k\left[{\textstyle\frac{3}{2}}\,y_{17}-3\,y_{18}+y_{19}\right] +
    f_1\left[y_{11}+y_{18}\right] + F_{R_y} y_5
    + {\rm i}\,R_x\left[-{\textstyle\frac{1}{2}}\,\phi_2 y_{17}+g_1 y_{11}+
      g_2 y_5\right], \nonumber \\
  y_{19}^{\prime} & = &
    k\left[-3\,y_{19}+y_{20}\right] +
    f_1\left[y_{12}+2\,y_{17}+y_{19}\right] + F_{R_y} y_6
    + {\rm i}\,R_x\left[\phi_1 y_{17}+g_1 y_{12}+g_2 y_6\right],
    \nonumber \\
  y_{20}^{\prime} & = & -3k y_{20} + f_1\left[y_{14}+y_{20}\right] -
    f_2 y_{17} +
    F_{R_y} y_8
    + {\rm i}\,R_x\left[-\phi_2 y_{17} + g_1 y_{14} + g_2 y_8\right].
    \nonumber \\
  & & \label{3ODE}
\end{eqnarray}
These equations have to be supplemented with the initial conditions
required for their numerical integration from $z=0$ to $z=1$:
\begin{equation}
  y_{20}\!\left(0\right) = 1,\quad y_i\!\left(0\right) = 0
  \quad\mbox{for}\quad i=1,\ldots,19. \label{3INI}
\end{equation}
In addition, we merely have to satisfy the fairly simple boundary condition
\begin{equation}
  y_1\!\left(1\right) = 0. \label{3BND}
\end{equation}

What matters here is precisely the replacement of the rather troublesome,
symmetric boundary conditions~(\ref{3SL2}) accompanying the original
Sturm--Liouville problem by the asymmetric boundary conditions (\ref{3INI})
and (\ref{3BND}), since these new boundary conditions lend themselves to a
standard shooting method: just integrate the system~(\ref{3ODE}) with the
initial conditions~(\ref{3INI}), monitor the resulting value of
$y_1\!\left(1\right)$, and adjust the system's parameters such that
this value becomes zero.

The price to pay for this simplification is a rather complicated system of
equations. But still, the spirit underlying these equations is simple. Since
we are not interested in the full spectrum of the eigenvalue
problem~(\ref{2SPC}) (or, equivalently, in that of the Sturm--Liouville
problem (\ref{3SL1}, \ref{3SL2})), but only in that value $R_c\argphi$ of $R$
where the lowest eigenvalue associated with some profile $\phi$ passes
through zero, we were entitled to set $\lambda = 0$ in the derivation of the
system~(\ref{3ODE}). We then keep both the profile $\phi$ and the wavevector
${\bf k}$ fixed when solving the initial value problem~(\ref{3ODE},
\ref{3INI}), and adjust only the parameter $R$. The smallest value of $R$
where $y_1\!\left(1\right)$ vanishes then equals that value of $R$ where the
smallest eigenvalue $\lambda$ becomes zero, for this particular $\phi$ and
${\bf k}$. We denote this $R$ as
\begin{equation}
  \Rzpk. \label{3R0P}
\end{equation}
The desired number $R_c\argphi$ then is identified as the global minimum,
taken over all wavevectors. Since the wavenumbers $k_x$ and $k_y$ effectively
enter into the system~(\ref{3ODE}) only through their squares $k_x^2$ and
$k_y^2$, we have
\begin{equation}
  R_c\argphi=
  \min_{k_x\geq 0,\,k_y>0}\left\{\Rzpk\right\}.
\end{equation}
Thus, the evaluation of the spectral constraint, i.e., the determination of
the maximal rescaled Reynolds number $R_c\argphi$ up to which a given profile
$\phi$ remains an admissible test profile for the variational
principle~(\ref{2VAP}), requires some labor. One first has to determine, for
that $\phi$ and a representative set of wavevectors ${\bf k}$, the first zero
of $y_1\!\left(1\right)$ that occurs when $R$ is increased from zero to
positive numbers. This yields the values~(\ref{3R0P}). In a second step one
has to locate the global minimum of all these values, which equals
$R_c\argphi$.

\section{Results}
\label{S_RE}

Now that we are in a position to evaluate the spectral constraint (\ref{2SPC})
for an arbitrary profile $\phi\argz$, we have to specify a class of test
profiles that is likely to exhaust the variational principle (\ref{2VAP},
\ref{2CEP}). Based on the analysis of the restricted Couette problem, we
propose the following class of variational profiles:
\begin{equation}
  \phi\argz = \left\{
  \begin{array}{ll}
    \frac{1}{2}\left(1-p\right) + p\,z - \frac{1}{2}\left(1-p\right)
      \left(1-z/\delta\right)^n & \mbox{for}
      \quad 0\leq z\leq\delta \\
    \frac{1}{2}\left(1-p\right) + p\,z & \mbox{for}
      \quad \delta<z<1-\delta \\
    \frac{1}{2}\left(1-p\right) + p\,z + \frac{1}{2}\left(1-p\right)
      \left(1-(1-z)/\delta\right)^n & \mbox{for}
      \quad 1-\delta\leq z\leq 1
  \label{4PRO}
  \end{array}\right..
\end{equation}
These profiles contain three independent parameters:
\renewcommand{\labelenumi}{\roman{enumi}.)}
\begin{enumerate}
  \item the boundary layer thickness $\delta$
    ($0<\delta\leq\frac{1}{2}$),
  \item the slope $p$ of the profile in the interior ($0<p\leq 1$),
    and
  \item the polynomial order $n$ ($n=4,5,\ldots$) of the boundary
    layer segments.
\end{enumerate}
By construction, each $\phi$ is $n-1$ times continuously differentiable at
the matching points $z=\delta$ and $z=1-\delta$. Since the right hand side
of the system (\ref{3ODE}) of differential equations contains derivatives
of the profile up to the fourth order, the smallest value of $n$ allowed by
this system is four. In the case $n = 4$ one encounters a discontinuity of
the fourth derivative. But as long as this discontinuity remains finite, it
poses no problem for the numerical integration. For practical computations
we restrict the parameter $n$ to values less than or equal to $1000$.

At this point it needs to be emphasized that the variational parameter $n$
plays an important role for the numerical solution of the variational
principle. If this parameter was artificially kept fixed, that is, if one
considered the profiles (\ref{4PRO}) with a constant parameter $n=4$, say,
then one would inevitably be confronted with serious divergences which are
hard to deal with numerically, and therefore would drastically reduce the
maximal attainable Reynolds numbers. The formal reasons for these
difficulties lie in the expressions $\phi^{(2)}\argz/\phi^{\prime}\argz$ and
$\phi^{(3)}\argz/\phi^{\prime}\argz$, which enter into the system (\ref{3ODE})
through the auxiliary functions (\ref{3AUX}). Namely, for high Reynolds
numbers the optimized profiles tend to become flat in the interior, so that
one has to face small denominators.

Since the solution to this technical problem, achieved with the help of
the variational parameter $n$, is of outstanding practical importance,
we illustrate it with an example. Let us first note that the parameters
$\delta$ and $n$ are related to each other by the profile's slope at
$z=0$ (or, equivalently, at $z=1$):
\begin{equation}
  \phi^{\prime}\!\left(0\right)=p+\frac{n}{2\delta}\left(1-p\right).
  \label{4PP0}
\end{equation}
Secondly, we anticipate the high-$\Rey$-scaling behavior of the optimized
profiles: we have
\[
  \phi^{\prime}\!\left(0\right)\sim\alpha\Rey\quad\mbox{and}\quad
  \phi^{\prime}\!\left({\textstyle\frac{1}{2}}\right)=p\sim\beta\Rey^{-1}
\]
with constants $\alpha$ and $\beta$, as will be shown later on
(cf.~Eqs.~(\ref{4PPB}) and (\ref{4PPM})). This implies that the ratio
$n/(2\delta)$ in Eq.~(\ref{4PP0}) necessarily has to increase about
linearly with $\Rey$ when $\Rey$ becomes large. We now focus on two
different parameter combinations: (a) $\delta=0.5$ and $n=1000$, and
(b) $\delta=0.002$ and $n=4$. In both cases the ratio $n/(2\delta)$ takes
on the value $1000$, and what matters now is the way the divergences of
$\phi^{(2)}\argz/\phi^{\prime}\argz$ and $\phi^{(3)}\argz/\phi^{\prime}\argz$
manifest themselves when the profile slope $p$ approaches zero. Case~(a) is
actually met in high-$\Rey$-solutions of the variational principle when
the shape of the profiles' boundary segments is allowed to vary, whereas
this shape is artificially kept fixed in case~(b). In Figs.~\ref{F_2}
and~\ref{F_3} we demonstrate for these two cases the response of
$\phi^{(2)}\argz/\phi^{\prime}\argz$ to the variation of $p$ from $0.1$
to smaller values, depicting only the relevant intervals of~$z$. In case~(a)
the decrease of $p$ merely entails a shift of the smooth ascend from the
(negative) value at $z=0$ to zero. In marked contrast, in case~(b) the same
decrease of $p$ produces a rather sharp singularity in the vicinity of
$z=\delta$. The expression $-\phi^{(3)}\argz/\phi^{\prime}\argz$ shows,
qualitatively, the same behavior, but even more pronounced. While the
functions shown in Fig.~\ref{F_2} are easy to handle, a singularity like the
one that develops in Fig.~\ref{F_3} would soon limit the accessible values of
$p$, and thus prevent us from reaching asymptotically high Reynolds numbers.

At this point, an important difference between the full three-dimensional
plane Couette problem and the restricted, two-dimensional problem without
spanwise degrees of freedom shows up. The cumbersome denominator
$\phi^{\prime}$, which does {\em not\/} appear in the restricted
case, emerges when one eliminates $v_x\argz$ from Eq.~(\ref{3VZ1}) in order to
derive the Sturm--Liouville equation~(\ref{3SL1}). In this way one arrives at
Eq.~(\ref{3VX2}), whereas in the restricted case ($k_y=0$) the condition
(\ref{3DIV}) of divergence-freeness yields an immediate connection between
$v_x$ and $v_z^{\prime}$. Here lies the reason why an asymptotic theory of
upper bounds for the full Couette problem, paralleling the one developed for
the restricted case in Ref.~\cite{NGH97bc}, seems to be out of reach.
Nevertheless, the model problem has led to the identification of the test
profiles~(\ref{4PRO}) as profiles which will provide the best possible bounds
on $\ceps$ in the limit of large Reynolds numbers.

In Fig.~\ref{F_4} we visualize a generic representative of the profile
family~(\ref{4PRO}). The profile functional~(\ref{2FUN}), which becomes a
function of $\delta$, $p$, and $n$ when applied to the class~(\ref{4PRO}),
now reads
\[
  D\argphi=\left[\frac{1}{2}\frac{n^2}{\left(2n-1\right)\delta}-1\right]
    \left(1-p\right)^2\equiv{\cal D}\!\left(\delta,p,n\right).
\]
By constructing the lower envelope of all graphs of the type shown in
Fig.~\ref{F_1}, that are produced in the $\left(\Rey,\ceps\right)$-plane by
the individual test profiles, we are able to compute the optimal upper
bound on $\ceps$ obtainable from the profiles~(\ref{4PRO}) for Reynolds
numbers up to $\Rey\approx 10^6$. Moreover, from the optimized profiles
we can extract the asymptotic scaling of the profile parameters.

In Fig.~\ref{F_5} we report our findings for the optimized upper bound
$\cbar\argRe$ on $\ceps\argRe$. Remarkably, this variational bound separates
{\em with a sharp kink\/} from the lower bound $\clow\argRe=\Rey^{-1}$ at the
energy stability limit, which is located at
\begin{equation}
  \RES=82.65014887(1),\quad
  \ceps\!\left(\RES\right)=\RES^{-1}=0.012099191758(2).
  \label{4RES}
\end{equation}
(The numbers in brackets denote the uncertainty of the last digit.) Up to
$\RES$ both bounds coincide, see Eq.~(\ref{2EST}). The kink of the upper bound
at $\RES$ is no artifact caused by our particular choice of test profiles
(\ref{4PRO}): even though some numerical improvement of the upper bound
$\cbar$ is possible in the regime of intermediate $\Rey$ (see below), one will
still obtain a bound that {\em increases\/} for Reynolds numbers slightly
above $\RES$. This behavior is related to the singularity of the balance
parameter $a$ at $\RES$~\cite{NGH97a,NGH97bc}, and finds its explanation in a
change of the role of the optimal background flow. Below $\RES$ the optimal
background flow coincides with the laminar flow, that is, with a stationary
solution to the Navier--Stokes equations, but above $\RES$ this is no longer
the case. Therefore, the deviations ${\bf v}\argxt$ from the background flow
below and above $\RES$ have a different character. For $\Rey<\RES$ these
deviations are fluctuations around the physically realized laminar flow.
Since they will decay at least exponentially in time~\cite{J76,DR81}, they
yield no contribution to $\varepsilon$. For $\Rey>\RES$ the deviations can no
longer be regarded as ``fluctuations'' in the usual sense, and do contribute
to $\varepsilon$.

The apparent slight non-smoothness of our bound $\cbar$ at $\Rey\approx 200$,
on the other hand, indeed is an artifact; it is caused by the restriction
of~$n$ to values no less than four. We have found empirically that the
component $k_x$ of the wavevector minimizing $\Rzpk$ always vanishes. With
this knowledge, the system~(\ref{3ODE}) can be simplified: the 20-component
complex system is converted into a 20-component real one. As a consequence,
the parameter value $n=3$ becomes admissible, which results in a noticeable
improvement of the upper bound in the intermediate range $\RES<\Rey<700$, as
shown in the inset of Fig.~\ref{F_5}.

For Reynolds numbers around $740$ the upper bound exhibits a pronounced
minimum, and then ascends to the asymptotic value
\begin{equation}
  \lim_{\Rti}\cbar\argRe=0.01087(1). \label{4ASY}
\end{equation}
This value lies slightly above --- but within the uncertainty span of ---
Busse's asymptotic result~(\ref{2CBU}). Thus, our variational bound
possesses the same asymptotic $\Rey^0$-scaling as the bound provided by
the Optimum Theory. In Appendix~\ref{A_2} we treat the variational
principle {\em without\/} taking into account the condition
$\bnabla\bcdot{\bf V}=0$ in Eqs.~(\ref{2SPC}). This toy problem provides an
illustrative example for our methods, and allows a direct comparison of
variational results obtained numerically with the help of the
profiles~(\ref{4PRO}) to the corresponding results derived by
Howard~\cite{H72,B96} in the framework of the Howard--Busse theory. Our
numerical value for the asymptotic bound pertaining to this simplified case,
$\lim_{\Rti}\cbar\argRe\approx 0.07071$, differs from Howard's analytical
value $9/128\approx 0.07031$ by merely $0.6\%$, even though our variational
profiles are specifically adapted to the full problem, and {\em not\/} to this
simplified case.

A key for understanding the overall behavior of the variational upper
bound lies in the expressions $\Rzpk$, considered for fixed profile
$\phi$ as functions of the two-dimensional wavevector~${\bf k}$. We have
already mentioned the observation that for each considered test profile
the absolute minimum of $\Rzpk$ was attained for $k_x=0$. This finding
may seem surprising on the first glance, but becomes at least plausible
with the help of the following argument: if one neglects the condition
$\bnabla\bcdot{\bf V}=0$ in Eq.~(\ref{2SPC}), the minimizing
${\bf k}$-vector is the zero-vector, as shown in Appendix~\ref{A_2}. Taking
into account $\bnabla\bcdot{\bf V}=0$ then enforces that the minimizing
${\bf k}$ becomes non-zero. But since there is no characteristic length-scale
in streamwise direction, the minimizing $k_x$ remains zero.

There is a further feature that strongly influences the $\Rey$-dependence
of the optimized profile parameters, and the scaling behavior of the
resulting upper bound $\cbar\argRe$: the single minimizing wavenumber $k_y$
that characterizes the variational solution at comparatively low $\Rey$
bifurcates to a pair of minimizing wavenumbers, $k_{y,1}$ and $k_{y,2}$, at
the Reynolds number
\begin{equation}
  \RBi\approx 460.
\end{equation}
The variational principle possesses the intriguing property that above
$\RBi$ both corresponding minimal values of $\Rzpk$, $R_1$ and $R_2$,
are locked to exactly the same value, i.e., {\em two eigenvalues of the
eigenvalue problem (\ref{2SPC}) pass through zero simultaneously\/}.
This degeneracy reflects the fact that, physically speaking, the sensitivity
of the variational problem to the behavior of the profiles in the boundary
layers separates from the sensitivity to the profiles in the interior.
A fairly precise mathematical formulation of this statement can be given
for the Couette problem without spanwise degrees of freedom~\cite{NGH97bc}.

The consequences of this bifurcation for the upper bound on $\ceps$ are
dramatic: {\em its approximate $\Rey^{-1/4}$-dependence that prevails for
$\Rey\leq\RBi$ changes to the asymptotic $\Rey^0$-behavior,\/} while the
bound passes through its global minimum
\begin{equation}
  \cbar_{\text\it min}\approx 0.00970\quad\mbox{at}\quad
  \Rey_{\text\it min}\approx 740. \label{4MIN}
\end{equation}
In order to convey some feeling for the way the variational principle
works, we draw $\Rzpk$ as function of both components of ${\bf k}$,
for the optimized profiles corresponding to the Reynolds numbers $\RES$
(Fig.~\ref{F_6}; cf.\ Eq.~(\ref{4RES})) and $\Rey_{\text\it min}$
(Fig.~\ref{F_7}; cf.\ Eq.~(\ref{4MIN})), respectively. Figure~\ref{F_8}
depicts the minimizing values of~$k_y$ that belong to the variational bound
displayed in Fig.~\ref{F_5}. Whereas the upper $k_y$-branch scales
proportionally to $\Rey$ in the limit of large $\Rey$, the lower branch
approaches a constant value. This value is {\em precisely\/} the one that
also corresponds to the energy stability limit:
\[
  \lim_{\Rti}\frac{k_{y,1}}{2\pi}=\frac{\kES}{2\pi}\approx 0.4960.
\]
This finding is rather unexpected, since we have the simple laminar profile at
the energy stability limit, and seemingly entirely different, non-linear
profiles with an almost flat interior part in the asymptotic regime. However,
the restricted model problem has shown that this interior part of the
asymptotic profiles can be related to the laminar profile, so that the
asymptotic identity of $k_{y,1}$ and $\kES$ can be proven analytically by
mapping the corresponding eigenvalue problems to each other~\cite{NGH97bc}.

It is now of considerable mathematical and physical interest to investigate
the $\Rey$-dependence of the optimized profile parameters. We have to keep in
mind that the background flow ${\bf U}\argx=U\phi\argz\hatx$ is, by
construction, {\em not\/} a horizontally-averaged mean flow. On the other
hand, the optimal profiles $\phi$ invite comparison with physically realized
flow profiles~\cite{K97}, even though such a comparison remains speculative.
Fig.~\ref{F_9} shows the metamorphosis of the optimal profiles with increasing
Reynolds number. By a close inspection of the data underlying this figure, we
can clearly distinguish four different regimes:
\begin{enumerate}
  \item For $\RES\leq\Rey\leq\Rey_1\approx 160$, a {\em deformation of the
    laminar profile\/} takes place. Whereas the thickness $\delta$ of the
    boundary layers remains fixed at $0.5$, and the polynomial order $n$ of
    the boundary segments remains at the smallest possible value $3$, the
    profile's slope $p$ in the interior decreases from $1$ to almost the
    limiting value $0$. In this regime the upper bound $\cbar$ reaches its
    maximum and starts to descend.
  \item The following regime, $\Rey_1\leq\Rey\leq\Rey_2\approx 670$, is
    characterized by the {\em development of boundary layers\/}. The
    parameter $\delta$ decreases to its minimal value of about $0.14$,
    with $n$ remaining fixed at $3$, while the slope $p$ increases again.
    In this regime the bifurcation of the minimizing $k_y$ occurs,
    which alters the previous approximate $\Rey^{-1/4}$-behavior of the upper
    bound.
  \item In the next regime, $\Rey_2\leq\Rey\leq\Rey_3\approx 1845$, we
    observe a dramatic change of the shape of the optimized profile's
    boundary segments. The parameter $n$ --- characterizing the shape of
    these segments by means of their polynomial order --- increases from
    the smallest possible value $n=3$ to $n=34$, thus steepening the profiles
    in the immediate vicinity of each boundary, thereby effectively generating
    new internal boundary layers within the original boundary layers.
    As a consequence of this {\em structural reorganization of the boundary
    layers\/}, the entire boundary segments, the extensions of which are
    quantified by $\delta$, start to reach deeper and deeper into the
    bulk and finally join again, $\delta$ taking on its maximal value $0.5$.
    In this regime, the upper bound passes through its minimum, see
    Eq.~(\ref{4MIN}), and then changes the sign of its curvature.
  \item For $\Rey$ above $\Rey_3$, we find simple scaling laws for the
    optimal profile parameters. These scaling laws give rise to a
    {\em power-law change of the profile\/} with increasing $\Rey$: the slope
    at the boundary $z=0$ (or at $z=1$) is given by
    \begin{equation}
      \phi^{\prime}\!\left(0\right)\sim\frac{n}{2\delta}\sim\alpha\Rey,
      \label{4PPB}
    \end{equation}
    while the slope at the midpoint $z=\frac{1}{2}$ becomes
    \begin{equation}
      \phi^{\prime}\!\left({\textstyle\frac{1}{2}}\right)=p\sim\beta\Rey^{-1},
      \label{4PPM}
    \end{equation}
    with constants $\alpha$ and $\beta$. Parallel to this power-law change
    of the profile, the upper bound on $\ceps$ ascends smoothly to its
    asymptotic value.
\end{enumerate}

If we had not been forced by practical numerical reasons to constrain the
shape parameter $n$ to values less than or equal to $1000$, we would have
preserved the asymptotic $\Rey^1$-scaling of $n$, while the parameter $\delta$
would have remained fixed at $0.5$. In our numerics the optimal $n$ becomes
$1000$ at $\Rey\approx 51880$, and keeping $n$ fixed at $n=1000$ for even
higher $\Rey$ instead of allowing it to increase further then forces $\delta$
to decrease, as follows from Eq.~(\ref{4PPB}). The ensuing increase of the
asymptotic value of the upper bound $\cbar$ (compared to the value that would
be obtainable if one allowed arbitrarily large $n$) is quite small, as can be
seen from the fact that $\cbar\!\left(51880\right)=0.010832(1)$ differs merely
by about $0.3\%$ from the asymptotic value (\ref{4ASY}).

\section{conclusions}
\label{S_CO}

Figure~\ref{F_10} gives a synopsis of rigorous upper bounds on the
dimensionless rate of energy dissipation~(\ref{2NDD}) in plane Couette flow
that have been found so far. The topmost three lines indicate results derived
with the help of an over-restrictive profile
constraint~\cite{DC92,GGHL95,NGH97a} stemming from functional estimates. This
sharpened constraint over-satisfies the actual spectral
constraint~(\ref{2SPC}) and thus leads to bounds exceeding the asymptotic
Busse bound~(\ref{2CBU}) by factors of 8.8, 7.9, or 6.6, respectively. Only
if the spectral constraint is properly implemented and evaluated, as done in
this work, one can obtain a rigorous bound that practically coincides with
the Busse bound in the asymptotic regime, thereby confirming both Busse's
pioneering work~\cite{B70,B78} and Kerswell's recent analysis~\cite{K97}.

However, it must be clearly recognized that much more has been achieved than
the confirmation of a 25-year old difficult result. Now a variational
principle is available that can not only produce asymptotic bounds of high
quality, but yields rigorous bounds of the same quality in the entire range
from low to asymptotically high Reynolds numbers. For the plane Couette
system discussed in this paper, the bound shows a remarkable structure,
notably a pronounced minimum followed by a $\Rey$-range between $1000$ and
$1800$ in which the bound's curvature changes its sign. This occurs at about
those Reynolds numbers where typical laboratory shear flows start to become
turbulent. It is tempting to speculate about this coincidence, but such
considerations are beyond the scope of the present work.

Of considerable interest is the mechanism that determines the variational
upper bound. As depicted in Fig.~\ref{F_8}, the minimizing wavenumber $k_y$
bifurcates at $\RBi\approx 460$, giving rise to one minimizing $k_y$-branch
that asymptotically approaches a constant, and to another one that scales
linearly with $\Rey$. Since both corresponding minima determine the solutions
to the variational principle by taking on the same value, as described in
Sec.~\ref{S_RE}, there are two characteristic lengths that enter into the
solution, one scaling with $\Rey^0$, the other with $\Rey^{-1}$. The analysis
of the restricted Couette problem~\cite{NGH97bc} has led to the conclusion
that the first, $\Rey$-independent minimum reflects the response of the
variational principle to the optimized profiles solely in the interior of the
fluid layer, whereas the second minimum reflects the response to solely the
profiles' boundary segments. Thus, the two lengths entering the variational
solutions can be identified as basically (i) the extension of the interior
flat part of the optimized profile, and (ii) the effective width of its
boundary segments. It is structural insight of this kind that will be required
for the formulation of a more refined variational principle that might
ultimately allow one to decide whether there are corrections to classical
scaling.

For the moment being, the comparison of our variational bound with the
experimental data shown in Fig.~10 demands some modesty. The bound clearly is
far from being sharp; the difference between the variational bound $\cbar$
and the corresponding data measured by Rei\-chardt~\cite{R59} for the plane
Couette flow or by Lathrop, Fineberg, and Swinney~\cite{LFS92} for the
small-gap Taylor--Couette system still spans an order of magnitude. In short,
one important step has been made, but there is still a long way to go.

\vspace{3mm}

\noindent {\bf Acknowledgements:} This work was supported by the Deutsche
Forschungsgemeinschaft via the Sonderforschungsbereich ``Nichtlineare
Dynamik'', SFB~185, and by the German-Israeli-Foundation (GIF).

\appendix
\section{Compound matrix method for the sixth-order Sturm--Liouville problem}
\label{A_1}

Keeping in mind that we are not interested in the calculation of the
entire spectrum furnished by the Sturm--Liouville eigenvalue problem
(\ref{3SL1}) and (\ref{3SL2}), but in the determination of that value
of $R$ where the smallest eigenvalue passes through zero, we can set
$\lambda = 0$ right from the outset. We thus obtain the new sixth-order
Sturm--Liouville problem
\begin{eqnarray}
  v_z^{(6)}-2\,\frac{\phi^{(2)}}{\phi^{\prime}}\,v_z^{(5)}-
  \left\{3\,k^2+\frac{\phi^{(3)}}{\phi^{\prime}}-
    2\left(\frac{\phi^{(2)}}{\phi^{\prime}}\right)^2\right\}v_z^{(4)}+
  \left\{4\,k^2\,\frac{\phi^{(2)}}{\phi^{\prime}}+
    {\rm i}\,k_x R\,\phi^{\prime}\right\}v_z^{(3)}
  & & \nonumber \\
  +\quad\left\{3\,k^4+2\,k^2\left[\frac{\phi^{(3)}}{\phi^{\prime}}-
    2\left(\frac{\phi^{(2)}}{\phi^{\prime}}\right)^2\right]+
    {\rm i}\,\frac{k_x}{2}\,R\,\phi^{(2)}\right\}v_z^{(2)} & & \nonumber \\
  -\quad\left\{2\,k^4\,\frac{\phi^{(2)}}{\phi^{\prime}}+
    {\rm i}\,k_x R\left[
      k^2\phi^{\prime}-\phi^{(3)}+
      \frac{\left(\phi^{(2)}\right)^2}
      {\phi^{\prime}}\right]\right\}v_z^{\prime}
  & & \nonumber \\
  -\quad\left\{k^6+k^4\left[\frac{\phi^{(3)}}{\phi^{\prime}}-
    2\left(\frac{\phi^{(2)}}{\phi^{\prime}}\right)^2\right]-
    \frac{1}{4}\left(k_y R\,\phi^{\prime}\right)^2+\qquad\qquad\right. & &
    \nonumber \\
    \left.{\rm i}\,\frac{k_x}{2}\,R\left[
      k^2\,\phi^{(2)}-\phi^{(4)}+
      3\,\frac{\phi^{(3)}\phi^{(2)}}{\phi^{\prime}}-
      2\,\frac{\left(\phi^{(2)}\right)^3}{\left(\phi^{\prime}\right)^2}
    \right]\right\}v_z & = & 0, \label{ASL1}
\end{eqnarray}
together with the boundary conditions
\begin{equation}
  \left.v_z\argz\right|_{z=0,1}=0,\quad
  \left.v_z^{\prime}\argz\right|_{z=0,1}=0,\quad\mbox{and}\quad
  \left.\left[v_z^{(2)}\argz-\frac{1}{2\,k^2}\,v_z^{(4)}\argz\right]
  \right|_{z=0,1}=0. \label{ASL2}
\end{equation}
The eigenvalues of this problem are those values of $R$ where both
Eqs.~(\ref{ASL1}) and (\ref{ASL2}) are satisfied.

In order to integrate Eq.~(\ref{ASL1}), one has to pose suitable initial
conditions at one of the boundaries, let us say $z=0$. We define the
six-vectors
\[
  {\bf V}_z\argz\equiv
  \left(v_z\argz,v_z^{\prime}\argz,
  v_z^{(2)}\argz-\frac{1}{2\,k^2}\,v_z^{(4)}\argz,v_z^{(3)}\argz,
  v_z^{(4)}\argz,v_z^{(5)}\argz\right)^T,
\]
and consider the three linearly independent fundamental solutions
${\bf V}_{z,\,1}\argz$, ${\bf V}_{z\,,2}\argz$, and ${\bf V}_{z,\,3}\argz$
to (\ref{ASL1}) that emerge from the initial values
\begin{equation}
  {\bf V}_{z,\,1}\!\left(0\right) = \left(0,0,0,1,0,0\right)^T,\quad
  {\bf V}_{z,\,2}\!\left(0\right) = \left(0,0,0,0,1,0\right)^T,
  \quad\mbox{and}\quad
  {\bf V}_{z,\,3}\!\left(0\right) = \left(0,0,0,0,0,1\right)^T, \label{AINI}
\end{equation}
respectively. By construction, each of these three solutions satisfies
the conditions~(\ref{ASL2}) at $z=0$, and the most general solution
obeying these one-sided b.c.\ is just a superposition of them. That solution
which also satisfies the conditions~(\ref{ASL2}) at the other boundary $z=1$
is then singled out by the requirement
\[
  \mbox{Det}\left(\begin{array}{ccc}
    v_{z,\,1}\!\left(1\right) & v_{z,\,2}\!\left(1\right) &
    v_{z,\,3}\!\left(1\right) \\
    v_{z,\,1}^{\prime}\!\left(1\right) & v_{z,\,2}^{\prime}\!\left(1\right) &
    v_{z,\,3}^{\prime}\!\left(1\right) \\
    v_{z,\,1}^{(2)}\!\left(1\right)-
    \frac{1}{2\,k^2}\,v_{z,\,1}^{(4)}\!\left(1\right) &
    v_{z,\,2}^{(2)}\!\left(1\right)-
    \frac{1}{2\,k^2}\,v_{z,\,2}^{(4)}\!\left(1\right) &
    v_{z,\,3}^{(2)}\!\left(1\right)-
    \frac{1}{2\,k^2}\,v_{z,\,3}^{(4)}\!\left(1\right)
    \end{array}\right) = 0.
\]
But from the numerical point of view the evaluation of this determinant,
which in most cases necessitates to subtract large numbers of almost equal
magnitude from each other, is rather inconvenient. An efficient and
numerically stable way to bypass this difficulty has become known as
{\em compound matrix method\/} in the literature, see, e.g., Ref.~\cite{S92}
for an introduction to the method, or Ref.~\cite{NGH97bc} for applications
to the restricted Couette problem. For applying this technique to the full
Couette problem studied in this paper, we define a new vector
${\bf y}\argz=\left(y_1\argz,\ldots,y_{20}\argz\right)^T$, the $20$
components of which are proportional to the $3\times 3$ minors of the
$6\times 3$ solution matrix that contains ${\bf V}_{z,\,1}$ as its first,
${\bf V}_{z,\,2}$ as its second, and ${\bf V}_{z,\,3}$ as its third column:
\begin{eqnarray}
  y_1 & \equiv & k^9 \sum_{\sigma}\mbox{sign}\!\left(\sigma\right)\,
  v_{z,\,\sigma\!\left(1\right)}v_{z,\,\sigma\!\left(2\right)}^{\prime}
  \left(v_{z,\,\sigma\!\left(3\right)}^{(2)}-
        \frac{1}{2\,k^2}v_{z,\,\sigma\!\left(3\right)}^{(4)}\right),
  \nonumber \\
  & \vdots & \label{AYDE} \\
  y_{20} & \equiv & \sum_{\sigma}\mbox{sign}\!\left(\sigma\right)\,
  v_{z,\,\sigma\!\left(1\right)}^{(3)} v_{z,\,\sigma\!\left(2\right)}^{(4)}
    v_{z,\,\sigma\!\left(3\right)}^{(5)}. \nonumber
\end{eqnarray}
The summations involve all permutations $\sigma$ of $1,2,3$. The powers of
$k$ multiplying each sum are determined in the following way: the expression
defining $y_{20}$ carries the highest total number of derivatives, namely
$12$; this expression is multiplied by $k^0$. Descending in the index
from $20$ to $1$, the number of derivatives is successively diminished;
each derivative less gives a factor of $k$ more. This guarantees two things:
firstly, the initial conditions (\ref{3INI}) are independent of $k$, and
secondly, all components have the same order of magnitude.

From Eq.~(\ref{ASL1}) one then obtains a system of first-order equations
that already closely resembles the system~(\ref{3ODE}) which we have studied
numerically, with the only difference that in each component~$i$
($i = 1,\ldots,20$) the term $-3\,k\,y_i$ appearing in~(\ref{3ODE}) is still
missing. It is obvious that the initial conditions (\ref{3INI}) and the
boundary condition (\ref{3BND}) follow directly from (\ref{AINI}) and the
definitions (\ref{AYDE}).

This term $-3\,k\,y_i$ in the $i$-th component of the system is of particular
importance. The restricted Couette problem has taught us the
lesson~\cite{NGH97bc} that the system of first-order equations directly
provided by the compound matrix method is not suited for numerical analysis in
the high-$\Rey$-regime, where one has to monitor large values of $k$. One
rather has to rescale each component of the system by a common exponential
damping factor. The reason for this rescaling is related to the fact that the
optimized profiles tend to become flat in the interior of the fluid layer when
$\Rey$ becomes high, while the extension of the boundary segments, where the
profiles have to bend in order to meet the boundary conditions
$\phi\!\left(0\right)=0$ and $\phi\!\left(1\right)=1$, tends to zero. This
motivates to neglect the profiles' boundary conditions altogether, and to
consider Eq.~(\ref{ASL1}) with a linear profile $\phi\argz=cz$, so that
$\phi^{\prime}$ becomes equal to the constant $c$, and all higher profile
derivatives vanish. In the limit $c \to 0$ one then arrives at
\begin{equation}
  v_z^{(6)}-3\,k^2 v_z^{(4)}+3\,k^4 v_z^{(2)}-k^6 v_z=0, \label{ASLC}
\end{equation}
with the b.c.~(\ref{ASL2}). Applying the compound matrix technique as
sketched above to this boundary value problem, one finds a $20$-component
system of first-order equations as in the case of a general $\phi$, but
now with constant coefficients. This system can be solved analytically by
standard means, but an accurate numerical solution for large $k$'s will be
possible only if the solutions do not grow (or decrease) exponentially. Thus,
the intended numerical approach forces us to rescale the system such that the
largest eigenvalue of the rescaled system's coefficient matrix is exactly
equal to zero. It turns out to be possible to determine the Jordan normal
form of the $20\times 20$-matrix resulting from Eq.~(\ref{ASLC}) analytically;
the largest eigenvalue is non-degenerate and equals $3k$. Hence, the
transformation
\begin{equation}
  \tilde{y}_1\argz\equiv y_1\argz{\rm e}^{-3kz},\quad\ldots\quad,\quad
  \tilde{y}_{20}\argz\equiv y_{20}\argz{\rm e}^{-3kz}, \label{ARSC}
\end{equation}
has the desired effect. Applying the very same transformation also to the
system given by the compound matrix method for the case of a general
candidate profile $\phi$ (and finally omitting the tilde-signs), changes that
system merely be adding $-3\,k\,y_i$ to the $i$-th component, and thus
produces our system~(\ref{3ODE}). We emphasize that the identification of the
proper exponential scaling factor, and the actual scaling
transformation~(\ref{ARSC}), is crucial for obtaining equations that remain
numerically stable even in the regime of asymptotically high Reynolds numbers.

A striking feature found in the study of the restricted Couette problem,
where the system corresponding to~(\ref{3ODE}) consists of merely six
equations~\cite{NGH97bc}, is the possibility to reduce that {\em complex\/}
$6$-component system to a {\em real\/} $6$-component system, thereby
halving the number of equations. This simplification could be obtained from
taking $\mbox{Im}\!\left(y_1\argz\right)=0$, for all $z$. Even though we have
strong numerical evidence that the imaginary part of the component $y_1\argz$
vanishes also in the present case, we did not succeed in deriving a reduced
system for arbitrary~$\phi$. Hence, we make a virtue of necessity and
keep the full system~(\ref{3ODE}), but exploit our insight for controlling the
accuracy of the numerical solutions by monitoring the magnitude of
$\mbox{Im}\!\left(y_1\argz\right)$.

\section{Neglecting $\bnabla\bcdot{\bf V}=0$}
\label{A_2}

In this appendix we study the variational problem for energy dissipation
in plane Couette flow without requiring that the solutions ${\bf V}\argx$ to
the eigenvalue problem (\ref{2SPC}) be divergence-free. Then the solutions to
this eigenvalue problem are determined in an enlarged space of functions, so
that the bound obtainable for this simplified problem can naturally not be as
good as the one reported in Sec.~\ref{S_RE}. However, the distinct value of
this toy model lies in two points. Firstly, we can illustrate, condensed
in a nutshell, the principles of both the background flow method itself and
the use of the compound matrix technique. Secondly, this model allows an
immediate quantitative comparison of our numerically calculated, asymptotic
dissipation bound with the corresponding value provided by the Howard--Busse
theory.

To begin with, the Eqs.~(\ref{3VX1})--(\ref{3DIV}) for the spectral
constraint's eigenvalue problem simplify enormously when the condition
$\bnabla\bcdot{\bf V}=0$ is skipped:
\begin{eqnarray}
  \lambda v_x & = &
    -2\left(\partial_z^2-k^2\right)v_x+R\,\phi^{\prime}v_z, \label{BVX1} \\
  \lambda v_y & = &
    -2\left(\partial_z^2-k^2\right)v_y, \label{BVY1} \\
  \lambda v_z & = &
    -2\left(\partial_z^2-k^2\right)v_z+R\,\phi^{\prime}v_x. \label{BVZ1}
\end{eqnarray}
Equation (\ref{BVY1}) separates from the others, and we immediately obtain
a solution that is compatible with the b.c.~(\ref{3BCN}) for each
$\lambda$ and $k$, namely $v_y\argz=0$. Introducing the new functions
$v\argz\equiv v_x\argz-v_z\argz$ and $w\argz\equiv v_x\argz+v_z\argz$,
Eqs.~(\ref{BVX1}) and~(\ref{BVZ1}) transform into
\begin{eqnarray*}
  \lambda v & = & -2\left(\partial_z^2-k^2\right)v-R\,\phi^{\prime}v, \\
  \lambda w & = & -2\left(\partial_z^2-k^2\right)w+R\,\phi^{\prime}w;
\end{eqnarray*}
the b.c.\ read $v\!\left(0\right)=w\!\left(0\right)=0$ and
$v\!\left(1\right)=w\!\left(1\right)=0$.

We now assume that the profile functions $\phi\argz$ satisfy, besides the
conditions posed in Eq.~(\ref{2PRO}), also the additional monotony condition
\begin{equation}
  \phi^{\prime}\argz\geq 0\quad\mbox{for}\quad 0\leq z\leq 1. \label{BPPC}
\end{equation}
Our test profiles~(\ref{4PRO}) obviously comply with this plausible
requirement. Hence, if the component $w\argz$ does not vanish identically,
the positive-definiteness of $-\partial_z^2$ will enforce $\lambda>0$ for
positive $R$. Consequently, for $\lambda=0$ and $R>0$ we have $w\argz=0$,
and are left with the second-order Sturm--Liouville boundary value problem
\begin{equation}
  v^{\prime\prime}-
  \left(k^2-{\textstyle\frac{1}{2}}R\,\phi^{\prime}\right)v=0,\quad
  v\!\left(0\right)=v\!\left(1\right)=0. \label{BSLB}
\end{equation}
In exact analogy to the full problem, the main task posed by the spectral
constraint is to calculate for given $\phi$ and $k\geq 0$ the smallest
$R$-value $R\argphi\!(k)$ where (\ref{BSLB}) is fulfilled, and then to
minimize over all $k$ in order to extract $R_c\argphi$. But now the
second-order differential equation~(\ref{BSLB}) shows that the {\em minimizing
$k$-value $k_c\argphi$ equals zero\/} for each $\phi\argz$. In the final
step, the optimal upper bound on $\ceps\argRe$ follows from
inequality~(\ref{2CEP}).

The laminar profile $\phi\argz=z$ yields the ``energy stability limit''
for our toy problem,
\[
  \RES=2\pi^2\approx 19.74,\quad
  \ceps\!\left(\RES\right)=1/2\pi^2\approx 0.05066.
\]
In order to compute $R_c\argphi$ for non-laminar profiles we now rewrite the
second-order equation~(\ref{BSLB}) as a system of first-order equations by
defining the vector
${\bf y}\argz\equiv\left(v\argz,v^{\prime}\argz\right)^T$.
In this way we get
\begin{equation}
  y_1^{\prime}=y_2,\quad
  y_2^{\prime}=-{\textstyle\frac{1}{2}}R\,\phi^{\prime}y_1, \label{BODE}
\end{equation}
where we have set $k=0$; initial conditions at $z=0$ and b.c.\ at $z=1$
are given by
\begin{equation}
  {\bf y}\!\left(0\right)=\left(0,1\right)^T,\quad
  y_1\!\left(1\right)=0. \label{BINB}
\end{equation}
This reformulation of the Sturm--Liouville problem (\ref{BSLB}) corresponds
precisely to the reformulation provided by the compound matrix method.
If one compares this simple boundary value problem (\ref{BODE}, \ref{BINB})
to the corresponding boundary value problem (\ref{3ODE})--(\ref{3BND}), one
gets a vivid impression of the complications introduced into the variational
principle by the condition $\bnabla\bcdot{\bf V}=0$.

Figure~\ref{F_11} shows the numerically computed variational upper bound on
$\ceps\argRe$ that results from inequality~(\ref{2VAP}) when
$\bnabla\bcdot{\bf V}=0$ is neglected, i.e., when the spectral constraint is
translated into Eqs.~(\ref{BODE}, \ref{BINB}), and the previous test
profiles~(\ref{4PRO}) are employed. For simplicity, we have fixed the
parameter $n$ to the value $n=2$. Interestingly, this bound exhibits the same
shape as the one that had been found analytically~\cite{NGH97a} when another
over-restrictive profile constraint stemming from functional estimates had
been used instead of the actual spectral constraint; that bound is indicated
by the long-dashed line in figure~\ref{F_10}. The best possible asymptotic
upper bound $\cbar\argRe$ for the toy model without $\bnabla\bcdot{\bf V}=0$
can be found for a non-integer value of $n$, namely $n_{\infty}\approx 3.40$:
\[
  \lim_{\Rti}\cbar\argRe\approx 0.07071.
\]
This value has to be compared to the bound calculated analytically
by Howard~\cite{H72,B96}, also without accounting for divergence-freeness:
\[
  \lim_{\Rti}\cbar\argRe=\frac{9}{128}\approx 0.07031.
\]
The work of Kerswell~\cite{K97} suggests that both the Howard--Busse theory
and the background flow method, when fully exhausted, give the same asymptotic
upper bound. Even though we have deliberately used test profiles which are
{\em not\/} specifically adapted to the present toy model, we have missed
Howard's bound by merely $0.57\%$.

\begin{figure}
  \caption[Fig~1]{Graph of an upper bound on $\ceps$ produced according to
    Eq.~(\ref{2CEP}) by the profile~$\phi$ shown in the inset.}
  \label{F_1}
\end{figure}

\begin{figure}
  \caption[Fig~2]{Behavior of $\phi^{(2)}\argz/\phi^{\prime}\argz$ in the
    interval $0\leq z\leq 0.015$ for $\delta=0.5$ and $n=1000$. From left
    to right: graphs that result when $p$ is successively diminished
    by factors of $1/10$ from $p=0.1$ to $p=10^{-6}$.}
  \label{F_2}
\end{figure}

\begin{figure}
  \caption[Fig~3]{Behavior of $\phi^{(2)}\argz/\phi^{\prime}\argz$ in the
    vicinity of $z=\delta$ for $\delta=0.002$ and $n=4$. From top to bottom:
    graphs that result when $p$ is successively diminished by factors of
    $1/10$ from $p=0.1$ to $p=10^{-6}$, cf.~Fig.~\ref{F_2}.}
  \label{F_3}
\end{figure}

\begin{figure}
  \caption[Fig~4]{Test profile $\phi\argz$ as given by Eq.~(\ref{4PRO}),
    for a generic parameter set ($n=4$, $\delta=0.3$, $p=0.1$).}
  \label{F_4}
\end{figure}

\begin{figure}
  \caption[Fig~5]{Bounds on $\ceps$ for the plane Couette flow. Points denote
    the variational upper bound $\cbar\argRe$ computed numerically with the
    test profiles~(\ref{4PRO}); the solid line on the left is the lower bound
    $\clow\argRe=\Rey^{-1}$. The asymptotic value of the upper bound,
    $\lim_{\Rti}\cbar\argRe=0.01087(1)$, lies slightly above, but within the
    uncertainty span of Busse's asymptotic result~(\ref{2CBU}). --- The inset
    depicts the improvement obtained by preassuming that the wavevector
    minimizing $\Rzpk$ has a vanishing component in streamwise direction.
    Solid line: lower bound $\clow\argRe$; dashed line: same upper bound as in
    the main figure; points: improvement of the bound obtained by admitting
    the parameter value $n=3$.}
  \label{F_5}
\end{figure}

\begin{figure}
  \caption[Fig~6]{Three-dimensional surface plot visualizing the
    ${\bf k}$-dependence of $\Rzpk$ for the laminar profile, which is the
    optimized profile for Reynolds numbers up to the energy stability limit
    $\RES$.}
  \label{F_6}
\end{figure}

\begin{figure}
  \caption[Fig~7]{Three-dimensional surface plot visualizing the
    ${\bf k}$-dependence of $\Rzpk$ for the optimized profile
    belonging to the absolute minimum of the upper bound on $\ceps$ at
    $\Rey=\Rey_{\text\it min}\approx 740$. Note the appearance of two
    distinct minima on the $k_y$-axis. Since the profile results from
    the variational principle, both minima have the same value.}
  \label{F_7}
\end{figure}

\begin{figure}
  \caption[Fig~8]{Minimizing wavenumber(s) $k_y$ corresponding to the
    upper bound on $\ceps$ displayed in Fig.~\ref{F_5}.}
  \label{F_8}
\end{figure}

\begin{figure}
  \caption[Fig~9]{Three-dimensional plot visualizing the metamorphosis
    of the optimized variational profiles with the increase of the Reynolds
    number. We have depicted the most important $\Rey$-interval on a
    logarithmic scale, beginning with the energy stability limit $\RES$ and
    ending in the scaling regime $\Rey>\Rey_3$.}
  \label{F_9}
\end{figure}

\begin{figure}
  \caption[Fig~10]{
    Comparison of bounds on $\ceps\!\left(\Rey\right)$ for the plane Couette
    flow.
    --- Solid slanted straight line:
      lower bound $\clow\argRe=\Rey^{-1}$.
    --- Topmost horizontal solid line:
      upper bound obtained by Doering and Constantin in Refs.~\cite{DC94,DC92}
      with the help of an over-restrictive profile constraint and piecewise
      linear profiles; $\cbar\argRe\approx 0.088$ for $\Rey>11.32$.
    --- Dashed-dotted line:
      improved bound on $\ceps\argRe$ derived by Gebhardt \etal\/ in
      Ref.~\cite{GGHL95} from the analytical solution of the
      Doering--Constantin principle with the over-restrictive constraint;
      $\cbar\argRe\approx 0.079$ for $\Rey>16.98$.
    --- Long-dashed line:
      further improvement due to the introduction of the balance
      parameter in Ref.~\cite{NGH97a}; $\cbar\argRe\rightarrow 0.066$.
      This bound is still calculated analytically utilizing the
      over-restrictive profile constraint.
    --- Heavy dots:
      upper bound for the plane Couette flow obtained in this work from
      the variational principle~(\ref{2VAP}) with the actual spectral
      constraint (\ref{2SPC}), cf.\ Fig.~\ref{F_5};
      $\cbar\argRe\rightarrow 0.01087(1)$.
      The variational profiles are given by Eq.~(\ref{4PRO}), including $n=3$
      as an admissible parameter.
    --- Joining dashed line:
      asymptotic upper bound~(\ref{2CBU}) derived by Busse in
      Refs.~\cite{B70,B78}; $\cbar\argRe\rightarrow 0.010(1)$. The
      shaded area denotes the estimated uncertainty of this bound.
    --- Triangles:
      experimental dissipation rates for the plane Couette flow
      measured by Rei\-chardt~\cite{R59}.
    --- Circles:
      experimental dissipation rates for the Taylor--Couette system with small
      gap as measured by Lathrop, Fineberg and Swinney~\cite{LFS92}.}
  \label{F_10}
\end{figure}

\begin{figure}
  \caption[Fig~11]{Bounds on $\ceps\argRe$ for the plane Couette flow obtained
    by neglecting the condition $\bnabla\bcdot{\bf V}=0$ in the eigenvalue
    equations~(\ref{2SPC}). Points denote the variational upper bound
    $\cbar\argRe$, computed numerically from the test profiles~(\ref{4PRO})
    with fixed parameter $n=2$; the solid line on the left is the lower bound
    $\clow\argRe=\Rey^{-1}$.}
  \label{F_11}
\end{figure}

\end{document}